# Electronic band structure of novel tetragonal BiOCuS as a parent phase for novel layered superconductors

I. R. Shein* and A. L. Ivanovskii

*Institute of Solid State Chemistry, Ural Branch, Russian Academy of Sciences, Pervomaiskaya St., 91, Ekaterinburg, 620990 Russia*

**A B S T R A C T**

Very recently, the tetragonal BiOCuS was synthesized and declared as a new superconducting system with Fe-oxypnictide - related structure. Here, based on first-principle FLAPW-GGA calculations, the structural parameters, electronic bands picture, density of states and electron density distribution for BiOCuS are investigated for the first time. Our results show that, as distinct from related metallic-like FeAs systems, BiOCuS phase behaves as an *ionic semiconductor* with the calculated indirect band gap at about 0.48 eV. The superconductivity for BiOCuS may be achieved exclusively by doping of this phase. Our preliminary results demonstrate that as a result of hole doping, the [CuS] blocks become conducting owing to mixed Cu 3d + S 3p bands located near the Fermi level. For the hole doped BiOCuS the Fermi surface adopts a quasi-two-dimensional character, similarly to FeAs SCs.

* Corresponding author.
*E-mail address:* shein@ihim.uran.ru (I.R. Shein).

## 1. Introduction

Following the discovery of superconductivity at $T_C \sim 26K$ in LaFeAsO(F) [1], spirited search has led to a broad family of FeAs-based superconductors (SCs), see reviews [2-4]. Among them, five main groups of related FeAs SCs, which are known as «1111», «122», «111», «32225» and «42226» materials, have been found to date. The parent phases for these groups of FeAs SCs are: for "1111" – *Ln*FeAsO and *B*FeAsF; for "122" - *B*Fe$_2$As$_2$, for "111" - *A*FeAs, for «32225» - $Sr_3Sc_2Fe_2As_2O_5$ and for «42226» - $Sr_4M_2Fe_2As_2O_6$, where *Ln, A, B* and *M* are rare earth, alkaline, alkaline earth and transition metals, respectively.

All these FeAs SCs adopt a quasi-two-dimensional crystal structure, where $[Fe_2As_2]$ blocks are separated either by atomic *A, B* sheets (for three-component "111" or "122" phases), or by $[LnO]$, $[BF]$ blocks (for four-component "1111" phases) or by more complex perovskite-like blocks for five-component «32225» and «42226» phases. For all these FeAs SCs, the electronic bands in the window around the Fermi level are formed mainly by the states of the $[Fe_2As_2]$ blocks and play the main role in superconductivity whereas the mentioned atomic sheets or blocks serve as "charge reservoirs", see [2-4].

Very recently, a new four-component phase BiOCuS has been successfully synthesized [5]. This phase was considered as an As-free analogue for the above "1111" FeAs SCs; and Cu- or oxygen non-stoichiometry for BiOCuS leads to the occurrence of superconductivity below $T_C \sim 5.8$ K in this material.

In this Communication, we present the results of FLAPW-GGA calculations, which are performed with the purpose to understand the peculiarities of the electronic band structure and intra-atomic bonding for the new layered phase BiOCuS.

## 2. Structural models and computational aspects

BiOCuS was considered in a tetragonal unit cell, ZrCuSiAs type, space group P4/*nmm*, Z = 2, where blocks [BiO] are sandwiched with [CuS] blocks as depicted in Fig. 1. As the detailed atomic coordinates for BiOCuS are unknown, at the first stage the full structural optimization of this phase was performed both over the lattice parameters and the atomic positions including the internal coordinates $z_{Bi}$ and $z_S$. These self-consistent calculations were considered to be converged when the difference in the total energy of the crystal did not exceed 0.1 mRy and the difference in the total electronic charge did not exceed 0.001 $e$ as calculated at consecutive steps.

All our calculations were carried out by means of the full-potential method with mixed basis APW+lo (LAPW) implemented in the WIEN2k suite of programs [6]. The generalized gradient correction (GGA) to exchange-correlation potential in the PBE form [7] was used. The plane-wave expansion was taken to $R_{MT} \times K_{MAX}$ equal to 7, and the $k$ sampling with $13\times13\times5$ $k$-points in the Brillouin zone was used. Bi ($6s^26p^35d^0$), Cu ($3d^{10}4s^14p^0$), O ($2s^22p^4$) and S ($3p^23s^43d^0$) were treated as valence states. The hybridization effects were analyzed using the densities of states (DOSs), which were obtained by a modified tetrahedron method [8], and some peculiarities of the intra-atomic bonding picture are visualized by means of charge density map.

## 3. Results and discussion

### 3.1. Structural properties

The calculated lattice constants for BiOCuS ($a^{calc}$ = 3.9006 Å and $c^{calc}$ = 8.6481 Å) are in reasonable agreement with the available experiment ($a^{exp}$ = 3.8726 Å and $c^{exp}$ = 8.5878 Å [5]): the divergences $(a^{calc} - a^{exp})/a^{exp}$ and $(c^{calc} - c^{exp})/c^{exp}$ are 0.0072 and 0.0070, respectively. These divergences should be attributed to the presence of the secondary phases in the synthesized samples [5]. The calculated atomic positions are Bi: (¼; ¼; 0.1495), Cu: (¾; ¼; ½), S: (¼; ¼; 0.6640) and O: (¾; ¼; 0).

### 3.2. Electronic band structure

Figure 2 shows the band structure for BiOCuS phase with optimized geometry as calculated along the high-symmetry $k$ lines. Here, the main conclusion is that, in contrast to the related metallic-like "1111" FeAs systems, BiOCuS is a semiconductor with the calculated band gap (BG) at about 0.48 eV. Note that the valence band maximum is located between A and $\Gamma$ points, whereas the conduction band minimum is located at the Z point, indicating that this material belongs to indirect-transition type semiconductors. It is well-known that the BG values are underestimated within LDA-GGA based calculation methods usually by 30–70% [9-11]. For crude fitting of our LDA-GGA gap to the experimental value, using the correction factor 50%, we estimate the "experimental" BG for BiOCuS to be about 0.75 eV.

Figure 3 shows the total and atomic-resolved $l$-projected DOSs for BiOCuS. The quasi-core O 2s, S 3s and Bi 6s states (peaks A,B and C in Fig. 3) are located in the ranges from - 20 to -17.6 eV, from -14 to -12.5 eV and from -11.7 to -10 eV below the Fermi level $E_F$ = 0 eV and are separated from the near-Fermi valence band (peak D) by a gap. In turn, the valence states occupy energy interval from the Fermi level to -6.8 eV, and in this region the contributions from O 2p, S 3p, Cu 3d and Bi (s,p,d) states take place. Thus, the preliminary conclusion from our DOSs calculations is that the general bonding mechanism in BiOCuS is not of a simple ionic character, but includes also covalent interactions owing to the hybridization of mentioned valence states. Let us discuss the bonding picture for BiOCuS in greater detail.

*3.3. Chemical bonding.*

The covalent bonding character for BiOCuS phase may be well understood from the above site-projected DOSs calculations. As is shown in Fig. 3, Cu-S and Bi-O states are hybridized. These covalent bonds are clearly visible in Fig. 4, where the charge density map is depicted. On the other hand, the electron density between [CuS] and [BiO] blocks is practically absent, see Fig. 4; this implies that the ionic bonding takes place between these [CuS]/[BiO] blocks. To illustrate the nature of the mentioned ionic inter-blocks bonding, it is possible to use a simple ionic picture, which considers the usual oxidation numbers of atoms: $Bi^{3+}$, $Cu^{1+}$, $S^{2-}$ and $O^{2-}$. Thus, the charge states of the blocks are $[Bi^{3+}O^{2-}$ $]^{1+}$ and $[Cu^{1+}S^{2-}]^{1-}$, *i.e.* the charge transfer occurs from [BiO] to [CuS] blocks. Besides, inside these blocks, the ionic bonding takes place respectively between Bi-O and Cu-S.

Thus, summarizing the above arguments, the picture of chemical bonding for BiOCuS may be described as follows.

(i)  Inside [BiO] blocks, mixed ionic-covalent bonds Bi-O take place (owing to hybridization of valence states of Bi and oxygen atoms and Bi → O charge transfer);

(ii)  Inside [CuS] blocks, mixed ionic-covalent bonds Cu-S appear (owing to hybridization of valence states of Cu and S atoms and Cu → S charge transfer);

(iii)  Between the adjacent [BiO] and [CuS] layers, ionic bonds emerge owing to [BiO] → [CuS] charge transfer. Thus, generally the BiOCuS phase may be classified as an *ionic semiconductor*.

Note that as distinct from mixed metallic-ionic-covalent bonds for "1111'' FeAs SCs, for example, LaFeAsO, where Fe-Fe metallic-like bonds are due to the near-Fermi delocalized Fe $3d$ states, see [12], for BiOCuS the main bonds are of ionic and covalent types – as well as for other isostructural semiconductors:  LaZnAsO and YZnAsO [13].

*3.4. Doping effects.*

The above results show that the superconductivity for BiOCuS may be achieved *exclusively* as a result of doping of this phase. Really, the available experiments [5] show that the occurrence of the superconductivity for this material was observed for Cu- or oxygen deficient samples. From our calculations for ideal BiOCuS we can see an intense occupied peak (D', Fig. 3) near the band gap.

These data allow us to speculate that the metallic-like state with high density of states at the Fermi level $N(E_F)$ - as a factor for superconductivity for BiOCuS - may be achieved by partial "emptying" of the highest near-Fermi bands, *i.e.* by hole doping. Using the simple rigid-band model, we found that the Fermi level coincides with the maximum of peak D' by removal from the system at about 0.1 $e$.

Finally, for the mentioned "hole-doped" BiOCuS system the Fermi surface (FS) was obtained and depicted in Fig. 5. The results show that the FS consists of hole-like pockets along the direction Γ-Z, which adopt a quasi-two-dimensional-like topology, similar to those for similar to of "1111" FeAs SCs, see [2-4]. Moreover, the FS is formed mainly by the low-dispersive bands with Cu 3d + S 3p character from [CuS] blocks, which should be responsible for superconductivity for BiOCuS material.

## 4. Conclusions

In summary, by means of the FLAPW-GGA approach, we have studied the structural and electronic properties of the newly synthesized tetragonal quaternary phase BiOCuS, which was declared as a new superconducting system with Fe-oxypnictide - related structure [5].

Our band structure calculations show that the ideal stoichiometric BiOCuS behaves as *a semiconductor* with an indirect band gap. Using crude fitting of our LDA-GGA gap, we estimated the "experimental" BG for BiOCuS to be about 0.75 eV.

Our analysis reveals also that the intra-atomic bonding for BiOCuS has a *high-anisotropic character*, where mixed covalent-ionic bonds are inside [BiO] and [CuS] blocks,, whereas between the adjacent [BiO]/[CuS] blocks, ionic bonds emerge. Thus, the examined phase may be classified as an *ionic semiconductor.*

We assumed that the observed superconductivity for BiOCuS may be achieved *exclusively* by doping of this phase. Our preliminary data show that as a result of partial "emptying" of the valence band (*i.e.* by hole doping), it is possible to achieve the metallic-like state for BiOCuS with a high density of states at the Fermi level N($E_F$), which is a factor for superconductivity. In this case the Fermi surface adopts a quasi-two-dimensional-like topology, and is formed mainly by the low-dispersive bands from [CuS] blocks, which should be responsible for superconductivity for the BiOCuS material.

Naturally, further experimental and theoretical efforts are necessary to check up the above assumption and to develop the further doping strategy for BiOCuS based materials.

**Acknowledgments**

Financial support from the RFBR (Grant 09-03-00946-a) is gratefully acknowledged.

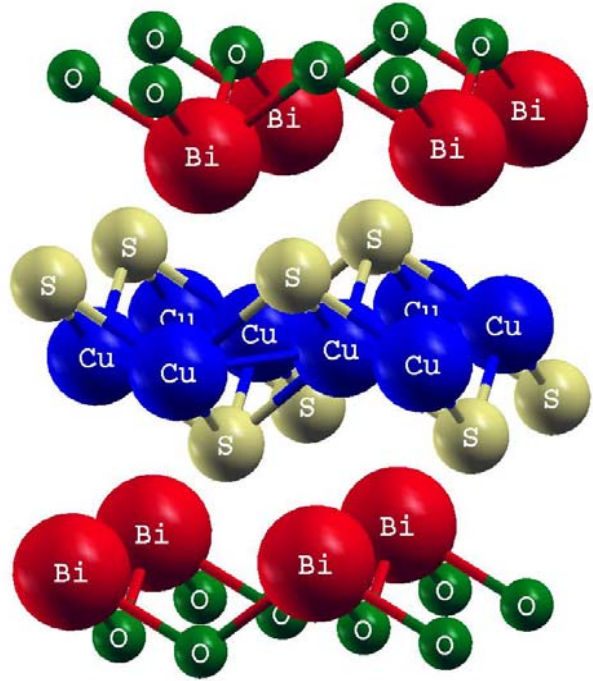

**Fig. 1**. Crystal structure of BiOCuS phase, space group *P4/nmm*.

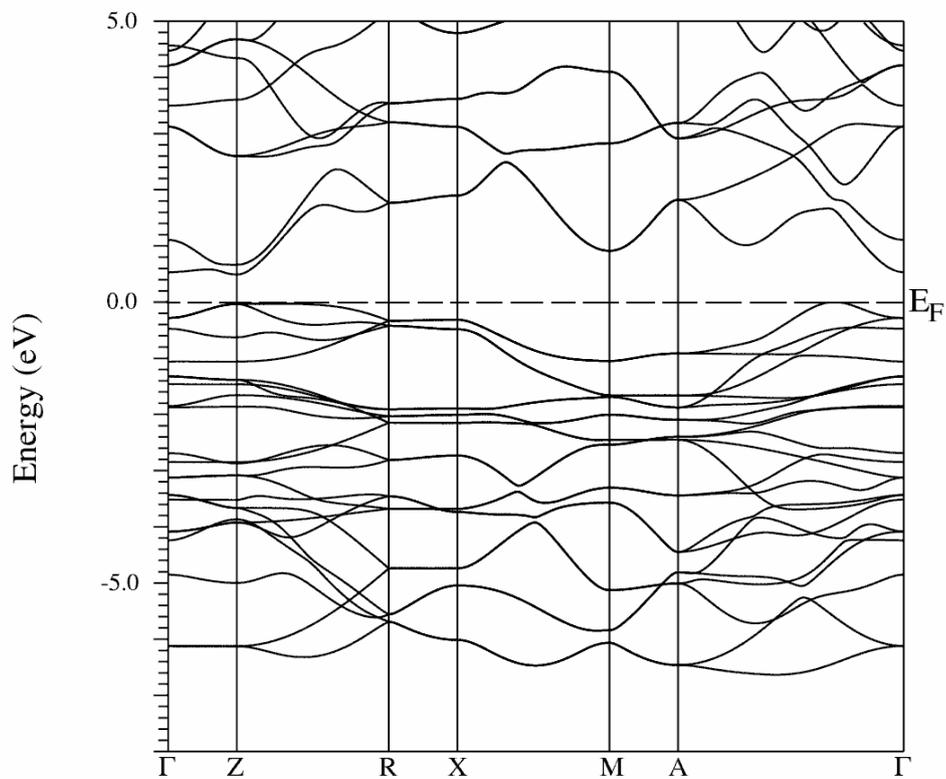

**Fig. 2**. Electronic bands for BiOCuS.

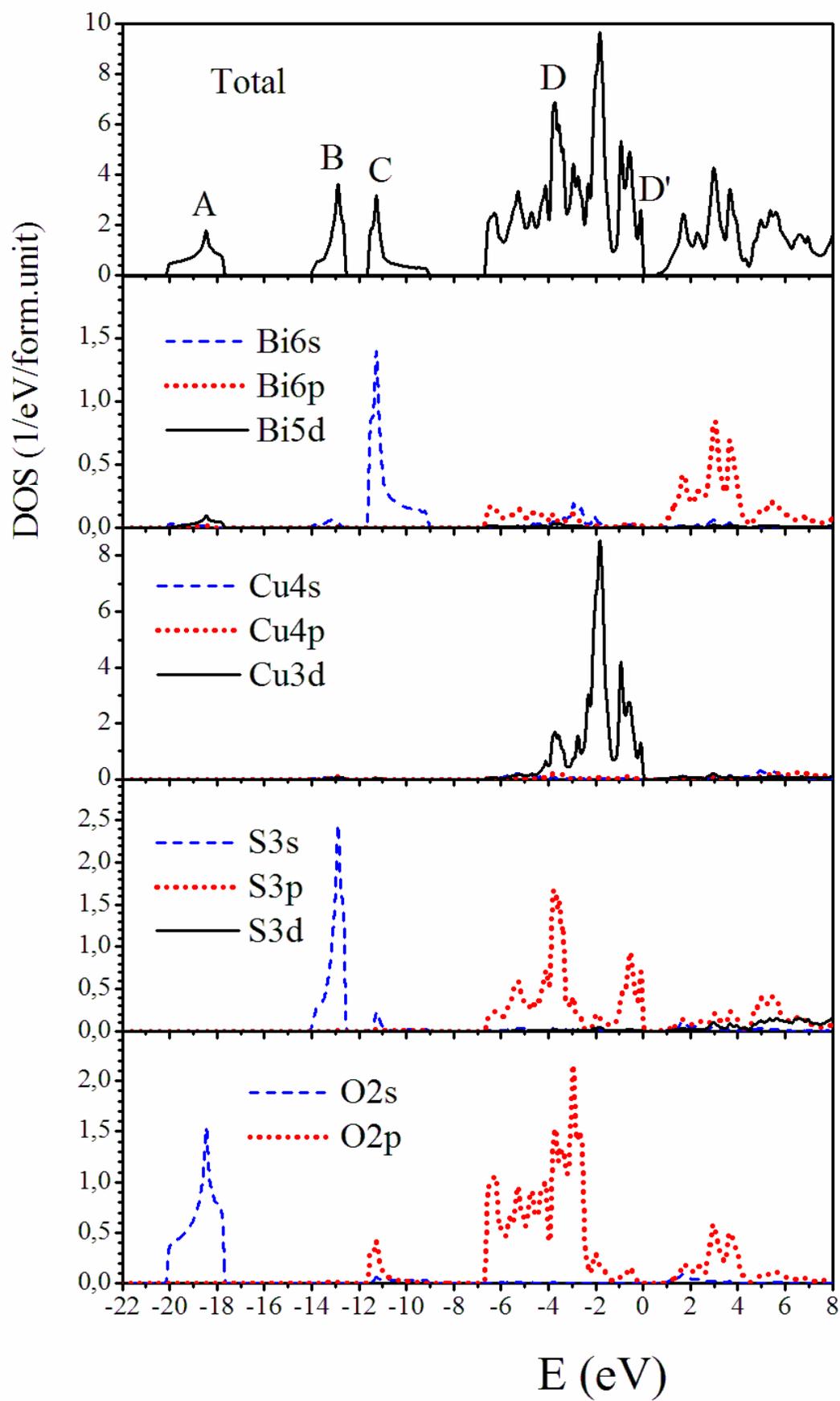

**Fig. 3**. Total (*upper panel*) and partial densities of states (*bottom panels*) for BiOCuS.

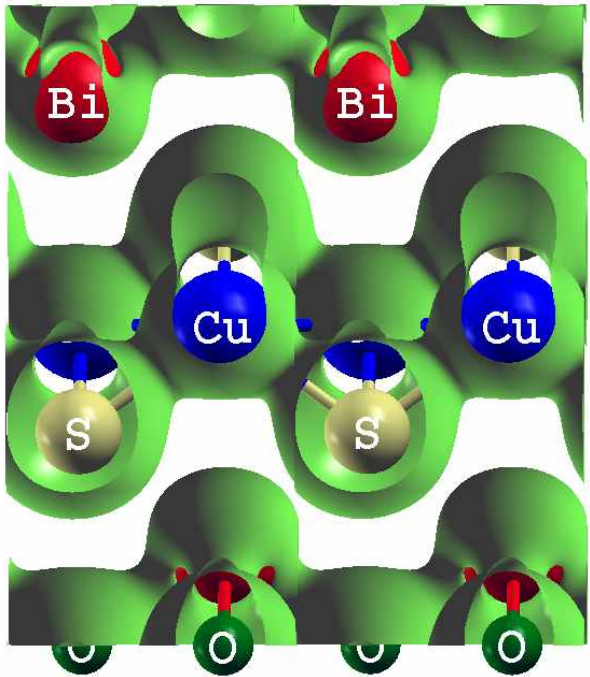

**Fig. 4**. Valence charge density map for BiOCuS.

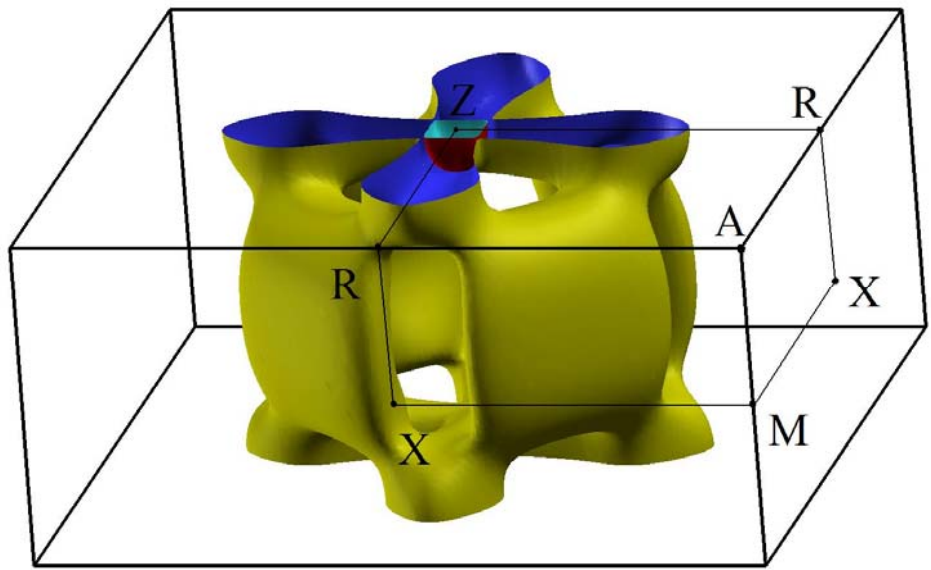

**Fig. 5**. Fermi surface for hole-doped BiOCuS (*see text*).